\documentclass[12pt,preprint]{aastex}

\usepackage{rotating}
\usepackage{amssymb}
\usepackage{longtable}
\usepackage{booktabs}
\usepackage{calc}

\shorttitle{Re-interpretation of SADs}
\shortauthors{Savage, McKenzie, \& Reeves}

\begin{document}

\title{Re-interpretation of Supra-Arcade Downflows in Solar Flares}

\author{$^{1}$Sabrina L. Savage, $^{2}$David E. McKenzie, \& $^{3}$Katharine K. Reeves}
\affil{$^{1}$NASA/Goddard Space Flight Center (Oak Ridge Associated Universities), 8800 Greenbelt Rd Code 671, Greenbelt, MD  20771, USA} 
\affil{$^{2}$Department of Physics, Montana State University, PO Box 173840, Bozeman, MT 59717-3840, USA}
\affil{$^{3}$Harvard-Smithsonian Center for Astrophysics, 60 Garden Street MS 58, Cambridge, MA  02138, USA}

\begin{abstract}

Following the eruption of a filament from a flaring active region, sunward-flowing voids are often seen above developing post-eruption arcades.  First discovered using the soft X-ray telescope aboard \textit{Yohkoh}, these supra-arcade downflows (SADs) are now an expected observation of extreme ultra-violet (EUV) and soft X-ray coronal imagers and spectrographs (e.g, \textit{TRACE}, \textit{SOHO}/SUMER, \textit{Hinode}/XRT, \textit{SDO}/AIA).  Observations made prior to the operation of AIA suggested that these plasma voids (which are seen in contrast to bright, high-temperature plasma associated with current sheets) are the cross-sections of evacuated flux tubes retracting from reconnection sites high in the corona.  The high temperature imaging afforded by AIA's 131, 94, and 193~$\mbox{\AA}$ channels coupled with the fast temporal cadence allows for unprecedented scrutiny of the voids.  For a flare occurring on 2011 October 22, we provide evidence suggesting that SADs, instead of being the cross-sections of relatively large, evacuated flux tubes, are actually wakes (i.e., trailing regions of low density) created by the retraction of much thinner tubes.  This re-interpretation is a significant shift in the fundamental understanding of SADs, as the features once thought to be identifiable as the shrinking loops themselves now appear to be ``side effects" of the passage of the loops through the supra-arcade plasma.  In light of the fact that previous measurements have attributed to the shrinking loops characteristics that may instead belong to their wakes, we discuss the implications of this new interpretation on previous parameter estimations, and on reconnection theory.

\end{abstract}


\section{\label{sads_20111022:intro}Introduction}

Magnetic reconnection is a ubiquitously occurring process in the universe and is widely accepted to be important for energy release in solar eruptions.  However, it is expected to occur in regions of low emission measure, and therefore, observations have tended to be indirect (e.g., \citeauthor{mckenzie_2002} \citeyear{mckenzie_2002}).  Retracting loops and sunward-flowing plasma voids above post-eruption flare arcades are often observed throughout long duration events using instruments sensitive to emissions from high temperature plasmas or white-light scattering from density structures in the corona (\textit{TRACE}, \textit{SOHO}/SUMER, \textit{SOHO}/LASCO, \textit{Hinode}/XRT, \textit{SDO}/AIA) (see \citeauthor{savage-mckenzie_2011}~\citeyear{savage-mckenzie_2011}, Figure~1 therein, for example images of these features).  Both supra-arcade downflowing loops (referred to as SADLs in \citeauthor{savageEA_2010}~\citeyear{savageEA_2010}, for example) and supra-arcade downflows (SADs: plasma voids) have been interpreted as the outflows created during the re-organization of the magnetic field during the reconnection process.  

To reconcile the difference in observational appearance between the plasma voids (SADs) and shrinking loops, a geometrical explanation was offered based on the line of sight to SADs or SADLs above curved polarity inversion lines (\citeauthor{savage-mckenzie_2011}~\citeyear{savage-mckenzie_2011}; \citeauthor{warren-obrien-sheeley_2011} \citeyear{warren-obrien-sheeley_2011}).  SADs were suggested to be the cross-sections of the retracting post-reconnection flux tubes, viewed along a line of sight tangential to their axes, while the shrinking loops were viewed from a line of sight orthogonal to their axes.  Many voids and shrinking loops have been reported in observations taken prior to the launch of \textit{SDO}/AIA (\citeauthor{mckenzie-hudson_1999}~\citeyear{mckenzie-hudson_1999}; \citeauthor{mckenzie_2000}~\citeyear{mckenzie_2000}; \citeauthor{innes-mckenzie-wang_2003a}~\citeyear{innes-mckenzie-wang_2003a}; \citeauthor{asaiEA_2004}~\citeyear{asaiEA_2004}; \citeauthor{sheeley-warren-wang_2004}~\citeyear{sheeley-warren-wang_2004}; \citeauthor{khan-bain-fletcher_2007}~\citeyear{khan-bain-fletcher_2007}; \citeauthor{reeves-seaton-forbes_2008}~\citeyear{reeves-seaton-forbes_2008}; \citeauthor{mckenzie-savage_2009}~\citeyear{mckenzie-savage_2009}; \citeauthor{savageEA_2010}~\citeyear{savageEA_2010}; \citeauthor{savage-mckenzie_2011}~\citeyear{savage-mckenzie_2011}).

\begin{figure}[!ht] 
\begin{center}

\framebox{
\includegraphics[width=.7\textwidth]{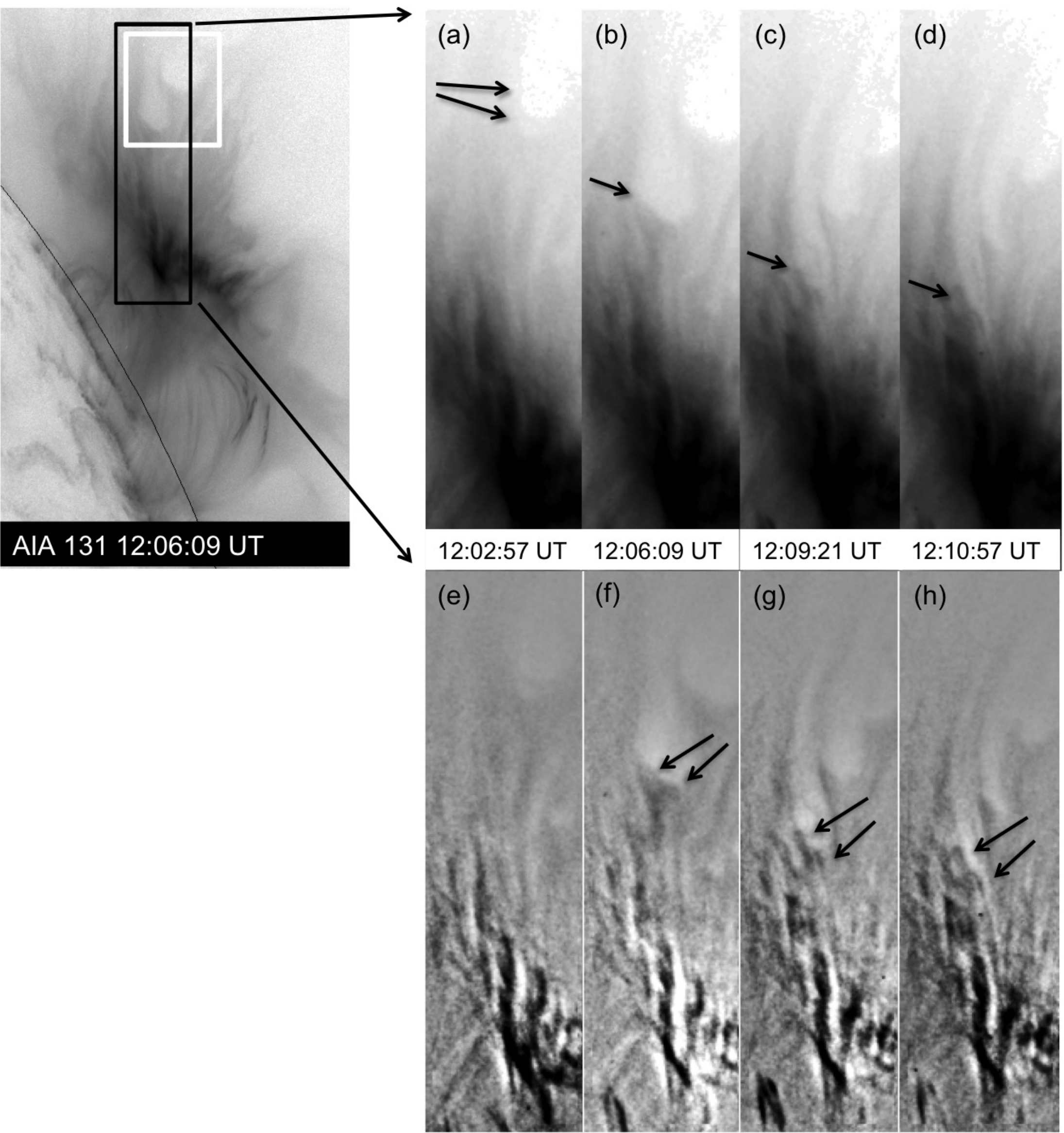}
}

\caption{Sequence of AIA 131~$\mbox{\AA}$ images (reverse-scaled) focusing on the evolution of one SAD.  The top panels show the descent of the leading edge of a large void.  Panel~(a) indicates the position of two initial SADs that appear to merge in subsequent frames.  Around 12:06~UT, two thin retracting loops appear at the leading edge of the void as indicated in the bottom panels, which have been differenced and scaled to enhance motion.  Panel (h) provides the clearest example image of the two leading loops. We refer the reader to the online movies accompanying this Letter for clearer evidence of this process.  The white box in the overlay panel contains two large SADs shown to be regions of density depletion in Figure~\ref{emmap.fig}.  (Animations of this figure are available in the online Astrophysical Journal.)}
\label{sad_raster}
\end{center}
\end{figure}

In this Letter, we present new evidence from recent AIA observations of the 2011 October 22 flare (SOL2011-10-22T10:00:00L045C065) that support an alternative interpretation of void-type SADs as \textit{wakes} behind thin retracting loops, rather than \textit{cross-sections} of much larger, evacuated loops.  For the purpose of this reference, we will refer to the density depletions that trail behind the shrinking loops as ``wakes" (i.e., disturbances in the density of the current sheet, apparently resulting from the passage of the shrinking loops).  The density within the thin loops is unknown due to their small size; however, the fact that they are emitting implies that they are not evacuated.  The key observation is depicted in Figure~\ref{sad_raster} and in the movies accompanying this Letter.

To be clear, SADs were also associated with the idea of a wake in the previous interpretation:  The shape of the plasma voids has been likened to that of a tadpole -- typically consisting of an oblong, tear-drop shaped trough at the head of the downflow followed by a thin, often waving ``tail" (\citeauthor{savage-mckenzie_2011}~\citeyear{savage-mckenzie_2011}, Figure~1 therein).  In the traditional interpretation, the head of the flow is the cross-section of a large evacuated flux tube, while the trailing tail has been explained as the wake of this large loop.  In the new interpretation, the tail is still considered part of the wake, but now the head of the flow is also \textit{completely} a wake and \textit{not} a loop cross-section.

An unresolved problem with the previous line-of-sight interpretation is the fact that the shrinking loop areas have been consistently measured to be much smaller than the void areas.  Loop areas measured with \textit{Yohkoh}/SXT, \textit{Hinode}/XRT, and TRACE range from $\sim$2 --~18~Mm$^{2}$ (median:  $\sim$2~Mm$^{2}$) versus the range for void areas of $\sim$2 -- 89~Mm$^{2}$ (median: $\sim$25~Mm$^{2}$) \citep{savage-mckenzie_2011}.  The new interpretation we present here would resolve the disparity in sizes between shrinking loops and downflowing plasma voids; however, these new findings call into question previously-reported parameter estimations such as fluxes and shrinkage energies.  

\clearpage

\section*{\label{sads_20111022:obs}Observations}

On 2011 October 22 at about 9:18 UT, \textit{SDO}/AIA \citep{lemenEA_2011} observed the slow eruption of a filament from AR 11314 (GOES M1.3) between $\sim$9:30 and 10:00~UT.  The eruption and arcade development were located between 10 and 20$^{\circ}$ in front of the west limb.   The passbands of interest for the observations reported herein are 131, 94, and 193~$\mbox{\AA}$ because the hot plasma in the supra-arcade region emits sufficiently at those wavelengths and the flows are clearly observable.   The plasma temperatures to which the narrow band AIA 193, 94, and 131~$\mbox{\AA}$ bandpasses admit significant response under flaring conditions are approximately 6~\&~20~MK, 7-10~MK, and 11-14~MK, respectively \citep{odwyerEA_2010}.  The angular resolution of the telescope is $\sim$0.6~arcseconds per pixel (corresponding to  $\sim$435~km).  AIA continuously observes the full Sun in all wavelengths at a 12~second cadence.

\begin{figure}[!ht] 
\begin{center}

\includegraphics[width=.95\textwidth]{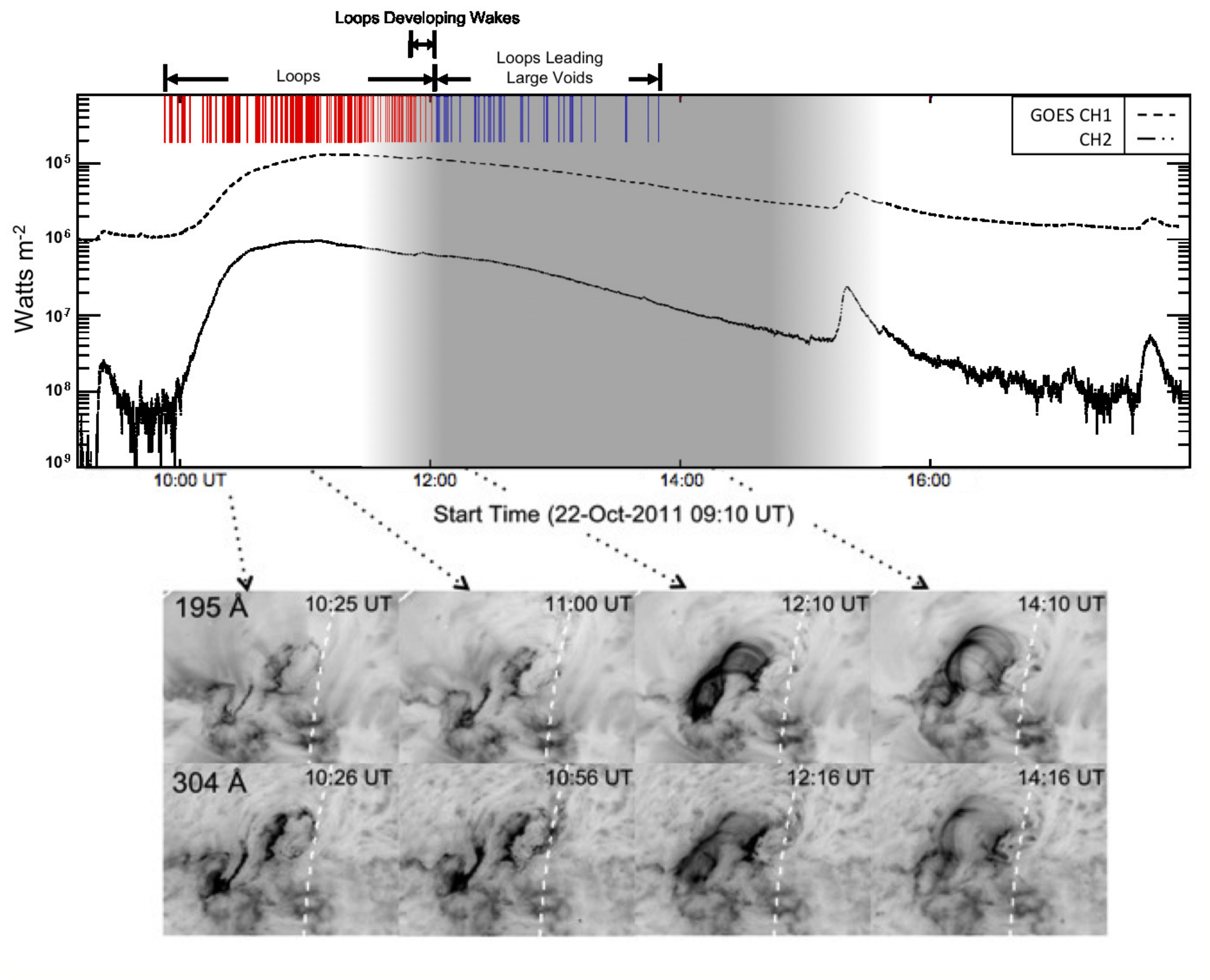}

\caption{\textit{Top}:  GOES light curve with downflow initial detection times overlaid at top (red: shrinking loops; blue: shrinking loops leading voids).  The shaded region indicates the presence of a fan of hot plasma in the supra-arcade region as seen primarily in the AIA~131, 94, and 335~$\mbox{\AA}$ channels.  The large peaks in the light curve after 14:00~UT are due to solar activity from other areas of the solar disk. \textit{Bottom}:  Sequence of \textit{STEREO-A}/SECCHI images taken with the 195 and 304~$\mbox{\AA}$ filters.  The image times per wavelength with respect to the GOES light curve are indicated by the arrows.  The white dashed curve represents the solar limb as seen from Earth's perspective.  A color version of this figure is available online.}
\label{goes_secchi_flows}
\end{center}
\end{figure}

The top panel of Figure~\ref{goes_secchi_flows} shows a GOES light curve with downflow initial detection times overlaid at top.  The downflows are classified as either shrinking loops only or loops leading large voids (the latter will be explained below).   The shaded region in Figure~\ref{goes_secchi_flows} indicates the presence of a fan of hot plasma in the supra-arcade region as seen in the AIA~131, 94, and 335~$\mbox{\AA}$ channels (and to a lesser degree in the 193~$\mbox{\AA}$ bandpass).  The shrinking loops are primarily observed prior to the presence of this plasma while the voids are only observed flowing through the fully-developed fan.  As the plasma is filling the region after $\sim$11:30~UT, wakes (for want of a better term) become marginally observable behind the shrinking loops.  If the wakes are present prior to this time period, they are undetectable.

The orientation of the footpoint ribbons and post-eruption arcade, as seen with the \textit{STEREO-A}/SECCHI 195 and 304~$\mbox{\AA}$ filters, is depicted in the bottom panel of Figure~\ref{goes_secchi_flows}.  The times for each wavelength panel with respect to the GOES light curve are indicated by the arrows.  This sequence of images shows that there is no significant change in the orientation of the arcade's axis throughout the event.   In addition, the arcade's axis appears to be simple and straight, not allowing for a single line of sight (as viewed from Earth) to be simultaneously tangent \textit{and} perpendicular to the polarity inversion line.  

\clearpage

\section{\label{sads_20111022:analysis}Analysis}

\begin{figure*}[!ht] 
\begin{center}

\includegraphics[width=.9\textwidth]{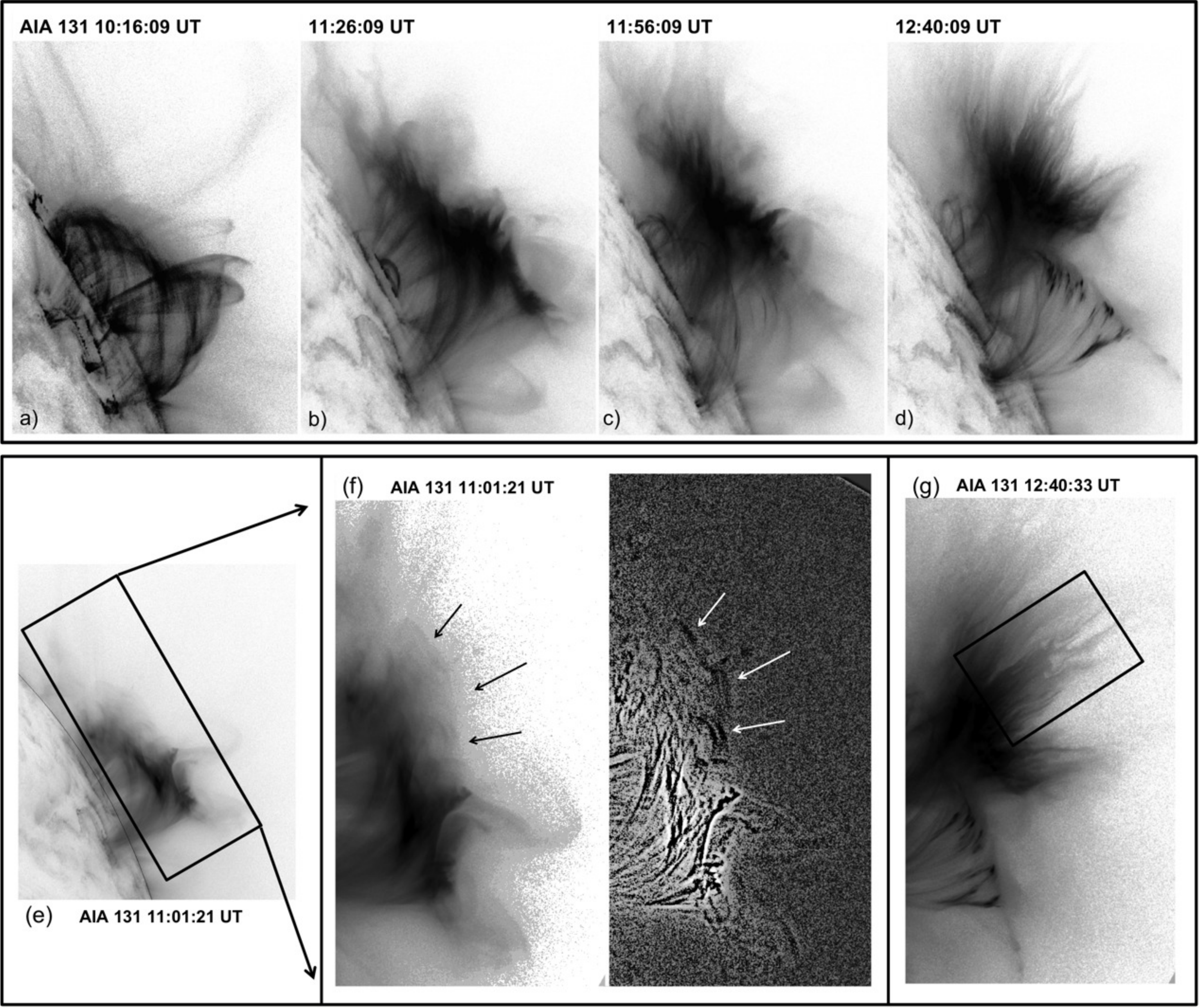}

\caption{\textit{Top}:  Sequence of AIA~131~$\mbox{\AA}$ showing the development of the arcade with no significant change in the orientation of the polarity inversion line.  \textit{Bottom}:  (e)  Image taken of the full active region with extractions (indicated by the box) shown in Panels (f) and (g).  (f)  Shrinking loops (indicated by the arrows) are observed during the early phase of the flare ($\sim$9:50$-$12:00~UT).  The left panel is the original image while the right has been differenced from the median of surrounding images and scaled to enhance motion.  (g)  SADs (contained within the boxed region) become apparent after about 12:00~UT.  All images are reverse-scaled.}
\label{sads_sadls_frames}
\end{center}
\end{figure*}

Beginning at $\sim$9:50~UT, descending loops are observed clearly in the 131 and 193~$\mbox{\AA}$ AIA channels.  Then near 12:00~UT, the predominant form of downflows shifts to plasma voids (SADs) without a change in viewing perspective (see Figures~\ref{goes_secchi_flows} and~\ref{sads_sadls_frames}).  The presence of shrinking loops and SADs in the same location along the same line of sight (as described in Section~\ref{sads_20111022:obs}) strongly indicates that viewing angle is not a differentiating factor between the two phenomena, as previously thought (\citeauthor{savage-mckenzie_2011}~\citeyear{savage-mckenzie_2011}; \citeauthor{warren-obrien-sheeley_2011} \citeyear{warren-obrien-sheeley_2011}).  Instead, the analysis indicates that it is the variation in the coronal density above the arcade that dictates the appearance of the downflows:  If the supra-arcade region is devoid of emitting plasma, the flows will be observed as shrinking loops.  If a hot fan of plasma is present above the arcade, the flows will appear as loops leading voids (SADs).

As done in \cite{mckenzie-savage_2009} and \cite{savage-mckenzie_2011}, the apexes of the shrinking loops were manually tracked while the plasma voids were tracked via a semi-automated trough-detection algorithm.  Typical measured speeds range between 50 and 500~km~s$^{-1}$, consistent with speeds previously reported (e.g., \citeauthor{savage-mckenzie_2011} \citeyear{savage-mckenzie_2011}; \citeauthor{warren-obrien-sheeley_2011} \citeyear{warren-obrien-sheeley_2011}, and references therein).  The sizes of the SADs range between $\sim$3 and 15~Mm$^{2}$, while the shrinking loops are nearer to $\sim$2 -- 3~Mm$^{2}$ prior to the development of the fan in the supra-arcade region.  These areas are also consistent with those previously reported using \textit{TRACE}, a similarly equipped instrument.  

The perspective of this flare, nearly perpendicular to the axis of the arcade as determined from analysis of the \textit{STEREO-A} data (Figure~\ref{goes_secchi_flows}), combined with AIA's high temperature coverage, resolution, and cadence, allows for further examination of the relationship between shrinking loops and SADs.  Figure~\ref{sad_raster} shows an image sequence of a large descending SAD led by shrinking loops.  The bottom set of images has been differenced from the median of surrounding images and scaled so as to accentuate the moving features.  There are two loops leading the void which may be due to the merger of two SADs creating the large descending void.  The arrows in the top sequence indicate the position of the SAD while those in the bottom sequence point out the loops as they become apparent at the leading edge of the void.  The positions of the two initial SADs are indicated in Panel~(a).  Admittedly, the lower SAD is much fainter and more difficult to distinguish in the single frame but is more readily identifiable in a high-cadence movie of the entire region.   (In order to enhance the visibility of the slow-moving loops, a subset of images with temporal spacing of 48~s between frames was selected.)  Panel~(h) provides the clearest example image of the two leading loops.  (We refer the reader to the online movies accompanying this Letter.)

The descending void shown in Figure~\ref{sad_raster} and the associated shrinking loops are observable between 11:58 and 12:45 UT beginning at a height of $\sim$150~Mm above the solar surface until the loops reach a potential configuration just above the cooler arcade ($\sim$60~Mm).  Most of the retraction occurs prior to 12:20~UT as the flows decelerate significantly near the arcade.  Even though SADs shrink considerably in size during their descent, this large void has a diameter that is much larger than the leading loops ($\sim$20~pixels [9~Mm] versus only $\sim$2$-$3~pixels [0.9$-$1.3~Mm]) even at its smallest near the top of the arcade (refer to Panel~(g)).  This analysis of this void reveals that SADs are actually the wakes of shrinking loops, rather than the loop cross sections.

While the images show that the largest voids appear to be wakes created by shrinking loops, the smallest voids are at the limit of AIA's resolution.  We consider it likely that they too are the wakes of loops.  The images reveal similar intensity enhancements, with linear extensions, at the leading edges of other SADs, though the loop-shaped morphology is difficult to resolve.

Consequently, instead of the observations shifting from shrinking loops to SADs, as noted at the start of Section~\ref{sads_20111022:analysis}, shrinking loops are always present.  The wakes, however, do not become apparent until there is a significant increase in hot material in the current sheet region.  A hot sheath surrounding the current sheet has been suggested by many previous observations (e.g., \citeauthor{svestka_1998} \citeyear{svestka_1998}; \citeauthor{gallagherEA_2002} \citeyear{gallagherEA_2002}), and is predicted in both analytical current sheet models \citep{seaton-forbes_2009} and numerical MHD flare/CME simulations (e.g., \citeauthor{reevesEA_2010} \citeyear{reevesEA_2010}; \citeauthor{yokoyama-shibata_2001} \citeyear{yokoyama-shibata_2001}) that include the effects of thermal conduction.

\begin{figure*}[!ht] 
\begin{center}

\includegraphics[height=200pt]{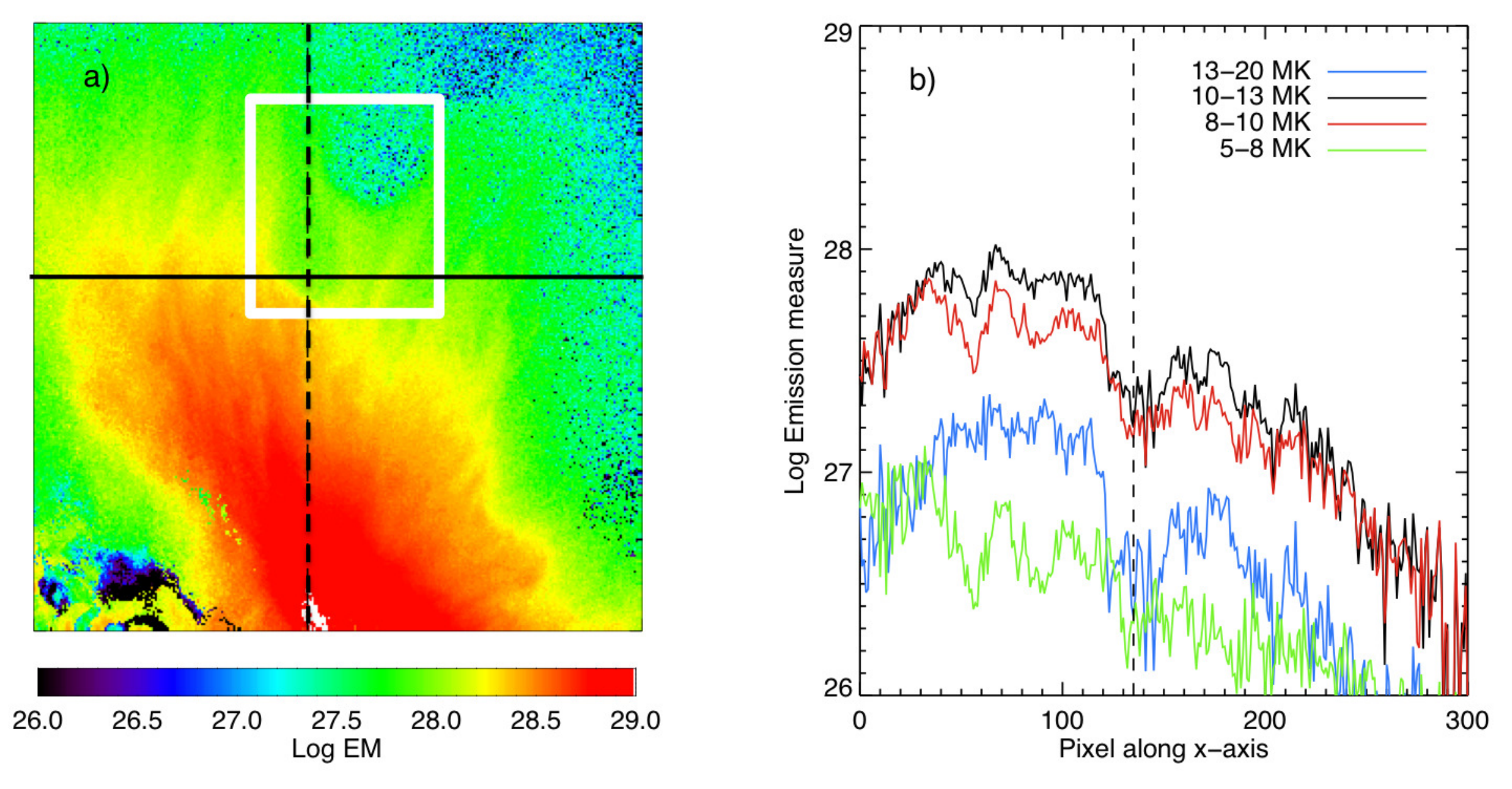}

\caption{Panel (a) shows a map, taken at 12:06:09~UT, of the emission measure at 10-13~MK derived from the DEMs calculated in each pixel.  Panel (b) shows a plot of the emission measure along the solid line plotted in Panel (a) for several temperature bands noted in the legend.  The vertical dotted line is in the same location in both the emission measure map and the plot, for reference.  The two large SADs contained within the white box are regions of density depletion.  These SADs are also highlighted in Figure~\ref{sad_raster} (overlay).}
\label{emmap.fig}
\end{center}
\end{figure*}

In order for the SADs truly to be density depletions in the current sheet, they must be shown to be regions of depressed density.  Spectroscopic observations of a previous flare have shown SADs to be voids \citep{innes-mckenzie-wang_2003a}.  There are no spatially resolved spectroscopic observations of the 2011 October 22 flare, so we use six AIA channels (94, 131, 171, 193, 211, \& 335~$\mbox{\AA}$) to construct differential emission measures to give an idea of the density within the voids.  We calculate differential emission measure curves (DEMs) in each AIA pixel using the six EUV images taken around 12:06:09~UT (the same time as Panel~(f) in Figure~\ref{sad_raster}).  We use the \verb"xrt_dem_iterative2" routine and the AIA response curves published in SolarSoft \citep{freeland-handy_1998}.  We estimate the error in the observations by using an empirical formula that approximates Poisson statistics for low count rates, but approaches Gaussian statistics for high count rates \citep[see, e.g.,][]{gehrels_1986}.

Figure \ref{emmap.fig} (a) shows an emission measure map constructed from the DEMs by integrating the DEM in each pixel over the 10-13~MK temperature range. Figure \ref{emmap.fig} (b) shows a plot of the emission measure for several temperature bins (5-8, 8-10, 10-13, and 13-20~MK) along a line that cuts through the SAD indicated with the arrows in Figure \ref{sad_raster}.  The emission measure in the SAD for the 10-13~MK temperature range is lower than the emission measure of the surrounding plasma by a factor of four.  Similar results are seen in a map of the emission measure at 8-10~MK.  In addition, there is no discernible increase across the SAD in the lower temperature bins, and there is no emission seen in the 1600 \& 304~$\mbox{\AA}$ images (which were not used to calculate the DEMs), indicating that there is not an incursion of cooler plasma.  The two large SADs in the upper part of the arcade, contained within the white box of Figure~\ref{emmap.fig} (a), also have emission measures that are lower than the surrounding plasma.  Thus, the SADs are indeed areas where the plasma density is depleted with respect to the rest of the supra-arcade plasma.  The density depletion of the SADs, along with their sizes and flow speeds, are characteristic of previously reported events.  


\newpage

\section{\label{sads_20111022:discussion}Discussion}

\begin{figure*}[!ht] 
\begin{center}

\includegraphics[width=.95\textwidth]{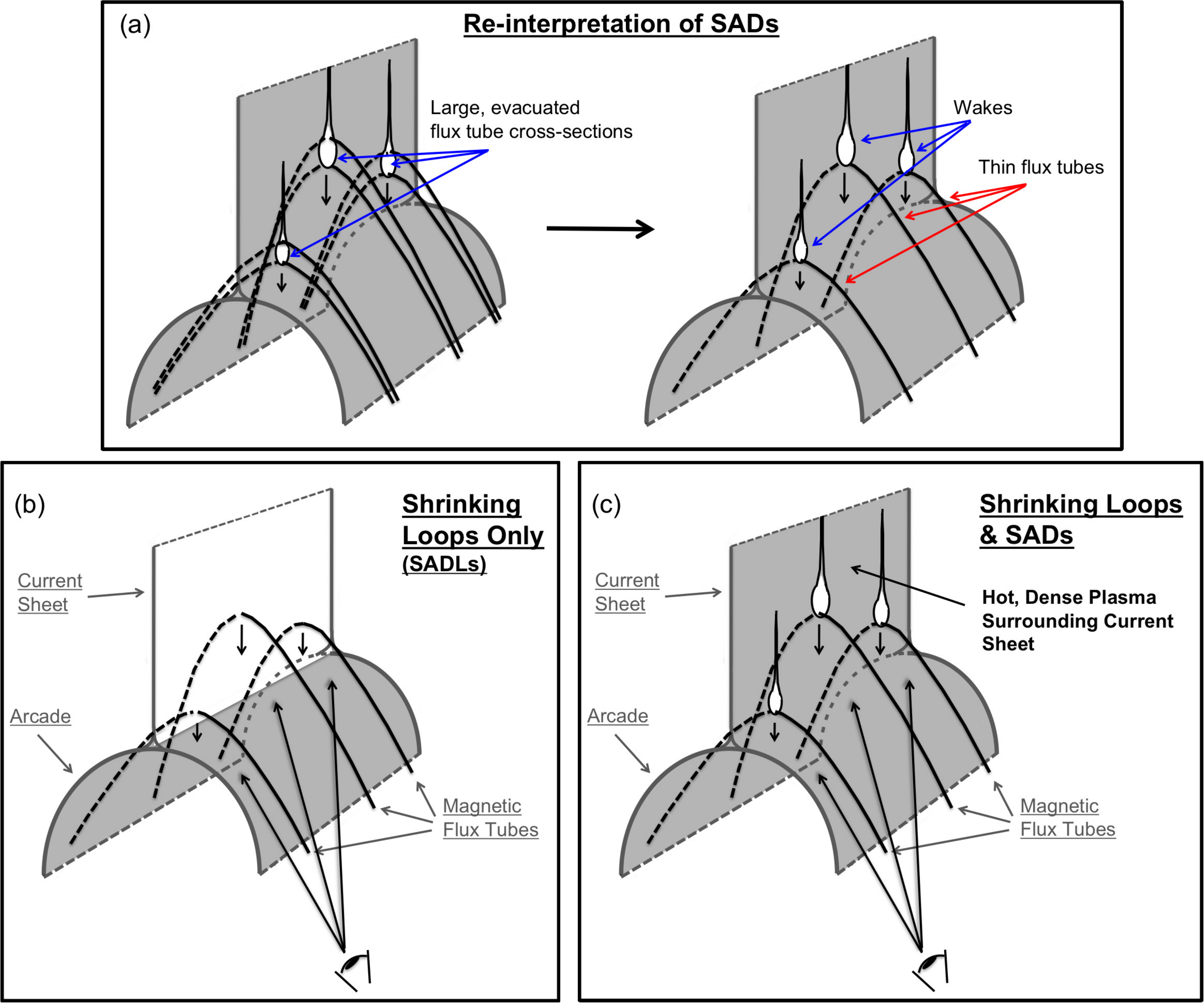}

\caption{(a)  Based on new evidence, the interpretation of SADs as the cross-sections of large, evacuated retracting flux tubes is no longer supported.  Rather, they appear to be the wakes of much thinner shrinking loops.  (b) Only shrinking loops are observed during the early phase of the eruptive event while SADs  become apparent after there is a significant increase in hot plasma (c), presumably surrounding the current sheet, in the supra-arcade region.  Note that the viewing orientation does not need to change to observe the two features; however, the increase in supra-arcade plasma is necessary to observe SADs (i.e., the wakes).  A color version of this figure is available online.}
\label{diagram_reinterpret_sads_sadls}
\end{center}
\end{figure*}

The increased temporal cadence, continual full disk field of view, and temperature coverage of the \textit{SDO}/AIA observatory has provided new information upon which to base the present re-interpretation of SADs.  Namely, SADs should no longer be considered cross-sections of newly-reconnected, large evacuated flux tubes; instead, these descending voids are apparently density depletions left in the wake of thin flux tubes retracting from a reconnection site in the supra-arcade region (Figure~\ref{diagram_reinterpret_sads_sadls}). 

There is the possibility of a high-beta plasma in the sheet, which is suggested by the observed vortical eddy-like motions, though concrete supporting evidence for high beta is still lacking.  Alternatively, the density depletions which we observe could be considered to result from the wave-interaction effects modeled by \cite{costaEA_2009}, \cite{schulzEA_2010}, and \cite{maglioneEA_2011}; or from plasma flows initiated/accelerated by the passage of shrinking loops.  Both of these effects are found in models of low-beta plasma.  The latter, field-aligned flows resulting in rarefaction behind the shrinking loops, is the subject of modeling currently underway.

In light of this development, previous parameter estimations of SAD characteristics require amending and re-evaluation.  For example, the sizes, fluxes, and shrinkage energy of post-reconnection flux tubes based on SADs measurements (\citeauthor{savage-mckenzie_2011} \citeyear{savage-mckenzie_2011}; \citeauthor{mckenzie-savage_2011} \citeyear{mckenzie-savage_2011}) are all over-estimates because the leading loop is much smaller in diameter than the void.  The relationship of a void area to the size of the associated flux tube, or to its flux, is not known at present.  However, we note that the smaller SADs tend to travel straight down into the arcade while the larger ones approach the arcade from an angle.

The presence of apparent wakes, and other complex flows revealed by the AIA data, are unexpected in the magnetically structured plasma of the supra-arcade region, and are not predicted in any current model of post-CME current sheet formation of which we are aware.  They raise questions about current sheet characteristics such as the plasma beta that surely must be investigated, but which are beyond the scope of the present Letter.  As aforementioned, magneto-hydrodynamic simulations of wake creation are underway.

While these observations indicate a fundamental difference from the original interpretation of SADs, the plasma voids are still inherently associated with reconnection outflows and thus remain probes of the reconnection process:  

1)  The presence of downflows in the flare decay phase illustrates that reconnection continues for many hours after the initial eruption.  Their presence in the impulsive phase, coinciding with non-thermal emissions (hard X-ray and microwave bursts), demonstrates that the SADs are associated with energy release (\citeauthor{khan-bain-fletcher_2007} \citeyear{khan-bain-fletcher_2007}; \citeauthor{asaiEA_2004} \citeyear{asaiEA_2004}).  

2)  The discreteness of SADs indicates that the reconnection is highly localized (i.e., ``patchy").  Similarly, temporal variation in the appearance of SADs indicates their production is bursty.  The reconnection in each location turns on and off on short timescales, independently of other locations \citep{linton-longcope_2006}.  

3)  The speeds of SADs and SADLs are approximately 50$-$500~km~s$^{-1}$, and are constant or slightly decelerating.  For comparison, typically assumed Alfv\'{e}n speeds are on the order of 1000~km~s$^{-1}$.  Reconnection models typically predict outflow speeds of (0.3$-$1.0)~$\times~v_{A}$ \citep{linton-longcope_2006}.  Deceleration may be expected, due to buildup of downstream magnetic pressure, though drag mechanisms may also be considered \citep{savage-mckenzie_2011}.

The revised relationship between SADs and shrinking loops is depicted in Figure~\ref{diagram_reinterpret_sads_sadls}.  The only distinguishing characteristic necessary between the two observational circumstances is the amount of hot plasma in the supra-arcade region surrounding the current sheet.  A change in orientation may indeed occur; however, SADs will \textit{only} be observed in the presence of this plasma.  

The sizes of the shrinking loops ($\sim$2$-$3 pixels [0.9$-$1.3~Mm]) are consistent with measurements for shrinking loops (i.e., SADLs) given in \cite{savage-mckenzie_2011}; however, many of the SADs are too small to detect a leading flux tube with \textit{SDO}/AIA, which is currently the highest resolution solar observatory capable of coronal measurements.  Therefore, the relevant size scales for post-eruption current sheet reconnection (i.e., the physical size of reconnection patches) appear to be less than $\sim$435~km (\textless~1~pixel).  

This long-duration event was also observed by RHESSI and \textit{SOHO}/LASCO; analysis of those data, which pertains to aspects of the flare other than the fundamental nature of the SADs, will be described separately in a forthcoming paper.


\section*{Acknowledgements}

S. L. Savage is supported by an appointment to the NASA Postdoctoral Program at Goddard Space Flight Center administered by Oakridge Associated Universities through a contract with NASA and under the mentorship of G. Holman.  D.E. McKenzie is supported under contract SP02H3901R from Lockheed-Martin to MSU.   K. K. Reeves is supported under contract SP02H1701R from Lockheed-Martin to SAO.

\end{document}